\newcounter{myctr}
\def\myitem{\refstepcounter{myctr}\bibfont\noindent\ifnum\themyctr>9\else\phantom{0}\fi\hangindent17pt\themyctr.\enskip}

\documentclass{ws-ijqi}

\newcommand{\ket}[1]{\left | #1 \right \rangle}
\newcommand{\bra}[1]{\left \langle #1   \right |}
\newcommand{\proj}[1]{\ket{#1}\bra{#1}}
\newcommand{\av}[1]{\langle #1\rangle}

\newcommand{\CC}{\mathbb{C}}

\newcommand{\Tr}{\operatorname{Tr}}

\newcommand{\comment}[1]{}
\newcommand{\ba}{\begin{eqnarray}}
\newcommand{\ea}{\end{eqnarray}}

\def\trimmarks{\relax}

\begin{document}

\markboth{L. P. Thinh, L. Sheridan and V. Scarani}
{Tomographic Quantum Cryptography Protocols are Reference Frame Independent}

\catchline{}{}{}{}{}

\title{TOMOGRAPHIC QUANTUM CRYPTOGRAPHY PROTOCOLS ARE REFERENCE FRAME INDEPENDENT}
\author{LE PHUC THINH, LANA SHERIDAN and VALERIO SCARANI}

\address{Centre for Quantum Technologies, National University of Singapore, 3 Science Drive 2, Singapore 117543\\Department of Physics, National University of Singapore, 2 Science Drive 3, Singapore 117542}

\maketitle

\begin{history}
\received{Day Month Year}
\revised{Day Month Year}
\end{history}

\begin{abstract}
We consider the class of \emph{reference frame independent} protocols in $d$ dimensions for quantum key distribution, in which Alice and Bob have one natural basis that is aligned and the rest of their measurement bases are unaligned.  We relate existing approaches to tomographically complete protocols.  We comment on two different approaches to finite key bounds in this setting, one direct and one using the entropic uncertainty relation and suggest that the existing finite key bounds can still be improved.
\end{abstract}

\keywords{quantum cryptography, quantum key distribution, reference frame independent,  uncertainty relations.}

\section{Introduction}
QKD protocols such as BB84\cite{bb84}, and six-state\cite{sixstate1}, as they are defined in the literature, require the alignment of Alice and Bob's local measurement frames: $X_A=X_B=X$, $Y_A=Y_B=Y$ and $Z_A=Z_B=Z$ where ${X,Y,Z}$ stand for the Pauli operators ${\sigma_X,\sigma_Y,\sigma_Z}$.  
It has been known for some time that alignment can be dispensed with, if one is ready to change the physical implementation of the qubit \cite{BGLPS04,TE11,AW07}; but these schemes require complex quantum states and are considered impractical.  More recently, it was shown that reference frame independence can be achieved by changing the protocol\cite{LSROB10}, while keeping the same physical implementation.  The main idea in the reference frame independent (rfi) protocol is that Alice and Bob share a common well-aligned measurement basis $Z=Z_A=Z_B$ (which is naturally available in many practical implementations~\footnote{Consider for examples, the time basis in a time bin implementation, the circular polarization basis, or the basis of different paths.}), while the other measurements can be misaligned by an arbitrary but fixed angle $\beta$
\begin{equation}
X_B=\cos\beta X_A+\sin\beta Y_A, \ \ \ Y_B=\cos\beta Y_A-\sin\beta X_A.
\end{equation}
Since the ``natural basis'' is automatically aligned, we have access to the quantum bit error rate in the $Z$ basis,
\begin{equation}
Q=\text{Pr(error)}=\text{Pr(a$\neq$b)}=\frac{1-\av{Z\otimes Z}}{2} .
\end{equation}
However, we also require additional information, namely the parameter $C$, in order to bound Eve's information.

\section{The parameter $C$}
The original rfi protocol introduced the $\beta$-independent parameter $C$,
\begin{equation}
\label{eq:C2}
C=\av{X_A\otimes X_B}^2+\av{X_A\otimes Y_B}^2+\av{Y_A\otimes X_B}^2+\av{Y_A\otimes Y_B}^2 , 
\end{equation}
which is an entanglement witness ($C\leq1$ for separable states and $C=2$ for maximally entangled states) to bound Eve's information.

We note that the protocol can be generalized to qudits in the following manner. Denoting $\{\ket{0},\ket{1},...,\ket{d-1}\}$ the computational basis vector of the Hilbert space describing a qudit, it is well known that the Pauli operators admit a generalization to higher dimension known as the Weyl operators, which are unitary operators of the form $X^kZ^\ell$ for $k,\ell\in\{0,1,...,d-1\}$ and 
\begin{equation}
Z=\sum_{j=0}^{d-1} \omega^j\proj{j}, \ \ X=\sum_{j=0}^{d-1}\ket{j+1}\bra{j},
\label{eq:def}
\end{equation}
where $\omega=e^{2\pi i/d}$ are the roots of unity and $j+1$ denotes the sum modulo $d$. To accommodate relative unitary rotation around $Z$, let $X_A=UXU^\dagger$ and $X_B=VXV^\dagger$ where $[U,Z]=[V,Z]=0$. In the protocol, Alice and Bob perform the projective measurements on the eigenstates of $X_A^{k_1}Z^{\ell_1}$ and $X_B^{k_2}Z^{\ell_2}$ and from the statistics estimate $Q$ and
\begin{equation}
C = \mathop{\sum_{k_1,k_2=1}^{d-1}}_{\ell_1,\ell_2=0} |\langle X_A^{k_1} Z^{\ell_1} \otimes X_B^{k_2} Z^{\ell_2} \rangle |^2
\label{eq:Cd}
\end{equation}
to bound Eve's information. In the appendix we prove that (\ref{eq:Cd}) is a generalization of (\ref{eq:C2}) with all the desired properties: it is independent of the local unitaries $U$ and $V$ mentioned above and is an entanglement witness ($C\leq(d-1)^2$ for separable states and $C=d(d-1)$ for maximally entangled states).

\section{The tomographic approach}  The way measurement results are used in the rfi protocol and its generalized version is not optimal, in the sense that the tomographic information deducible from the measurement statistics can be used directly, instead of via the parameter $C$.  Estimating $C$ requires the knowledge of $[d(d-1)]^2$ correlators $\langle X_A^{k_1} Z^{\ell_1} \otimes X_B^{k_2} Z^{\ell_2} \rangle$ which can alternatively also be used to completely specify the state as\cite{P77}
\begin{equation}
\rho_{AB}=\frac{1}{d^2}\mathop{\sum_{k_1,k_2,\ell_1,\ell_2=0}^{d-1}} \langle X_A^{k_1} Z^{\ell_1} \otimes X_B^{k_2} Z^{\ell_2} \rangle X_A^{k_1} Z^{\ell_1} \otimes X_B^{k_2} Z^{\ell_2}
\label{eq:rhoexpression}
\end{equation}
in the measurement bases of Alice $\{X_A^{k_1}Z^{\ell_1}\}$ and Bob $\{X_B^{k_1}Z^{\ell_1}\}$.  However, working only from $C$ discards some information that can lead to a tighter bound on Eve's information.  Using tomography, one can have a rfi protocol without the need for $C$\cite{LKEKO03}. 

Let us explain in detail how that is possible. The most direct approach is to make $d^2$ measurements $X^kZ^\ell$ on each subsystem (Alice has one, Bob the other) and combine the measurement outcomes to find each correlator directly. This is very inefficient because it requires $d^2$ different estimates to be made with good precision, which requires many copies of the state. (Recall that collective attacks are the optimal attack in general for Eve in this scenario~\cite{KGR05,RGK05}.)  Also it is unnecessary because many of the Weyl operators have the same set of eigenvectors ($Z^\ell$ for $\ell=0,...,d-1$ for instance); hence the measurement statistics of one can be used to calculate the average values of all the others.  In general, the minimum number of measurements needed to completely specify the state is still unknown. However, if $d$ is prime, one can reconstruct the state by making only $d+1$ measurements corresponding to $d+1$ mutually unbiased bases on each subsystem, say $\mathcal{B}=\{Z,X Z^\ell:\ell=0,...,d-1\}$. After the measurements, Alice and Bob can estimate their marginal probability distribution locally, and if they share the measurement outcomes, the joint probability distribution $p(a,b|A\otimes B)$ where $A,B\in\mathcal{B}$. It is well known that the eigenbasis of any $X^kZ^\ell$ is among the eigenbases of observables in $\mathcal{B}$; therefore from $p(a,b|A\otimes B)$ we can compute all the average values using
\begin{equation}
\langle X_A^{k_1} Z^{\ell_1} \otimes X_B^{k_2} Z^{\ell_2} \rangle=\sum_{a,b} \lambda_a\lambda_bp(a,b|A\otimes B),
\end{equation}
where $\lambda_a$ is the eigenvalue associated to the eigenvector representing outcome $a$ of $X_A^{k_1} Z^{\ell_1}$ and $A$ is the operator in $\mathcal{B}$ with the same eigenbasis as $X_A^{k_1}$, ditto for Bob.  
Hence a full state reconstruction is possible by~(\ref{eq:rhoexpression}).

Once $\rho_{AB}$ is found relative to the (partially) unaligned frames, it is possible to find local rotations $U_A, U_B$ such that $U_A\otimes U_B\rho_{AB}U_A^\dagger\otimes U_B^\dagger=\tilde{\rho}_{AB}$ where $\tilde{\rho}_{AB}$ is a diagonal matrix of the eigenvalues of $\rho_{AB}$. This procedure is always possible if Alice and Bob's marginals are random, and if this is not the case, Alice and Bob can do the randomization themselves. Then $\tilde{\rho}_{AB}$ is a mixture of $d$-dimensional Bell states (maximally entangled states) in the new bases effected by the local rotations.  Using the eigenvalues, $\boldsymbol{\lambda}$, of this state, the asymptotic rate will be\cite{SS10}
\begin{equation}
r_\infty = \log{d} - H(\boldsymbol{\lambda}),
\end{equation}
and the finite key bounds can also be calculated from those methods to give,
\begin{equation}
r_{N} =  \frac{n}{N} \left(\min_{E|\mathbf{P\pm\mu}} H(A|E) - \text{leak}_{\text{EC}}/n - \frac{2}{n} \log \frac{1}{\epsilon_{\text{PA}}}- (2\log d+3) \sqrt{\frac{\log(2/\bar{\epsilon})}{n}}\right) ,
\label{eq:rn}
\end{equation} 
where the security parameter is  
$
\epsilon = \epsilon_{\text{EC}}+\epsilon_{\text{PA}} + n_{\text{PE}}\epsilon_{\text{PE}} +\bar{\epsilon}  ,
$
and $\epsilon_{\text{EC}}$ is the probability of failure of the error correction step, $\epsilon_{\text{PA}}$ the probability of failure of the privacy amplification,  $\epsilon_\text{PE}$ is the probability that the estimate of any measured parameter $\mathbf{P}$ is outside a tolerated range $\mu$. The minimization of $H(A|E)$ is done over all attacks of Eve, denoted by $E$, compatible with observed parameters $\mathbf{P}$ within tolerated fluctuations $\mu$.

We also note that doing tomography it is possible to relax the assumption of the ``natural basis'', allowing Alice and Bob to have one axis which they know is close to being aligned but is not perfectly aligned.  This will lead to an increase in the error rate $Q$.  For example, for the case $d=2$ the $Z$ axes must not differ by more than $41.5^{\circ}$ for $Q$ to be small enough to have any hope of generating the shared randomness required to grow a secret key.  However, it is a natural assumption in many settings to take one basis to be aligned.

\section{Bounds from Uncertainty Relations}
Recently, a tighter finite-key bound for the Bennett-Brassard 1984 (BB84) protocol\cite{bb84} has been found by Tomamichel et. al.~\cite{TLGR11}.   For some time there has been the idea of viewing the security of QKD as reducing to suitable URs.  After a series of works \cite{K06,RB09,BCCRR10}, this program was brought to completion by Tomamichel and Renner~\cite{TR10}, who indeed provided a UR involving the quantity that captures Eve's uncertainty in QKD: the \textit{smooth min-entropy conditional on a quantum observer}. The result is directly applicable to finite-key bounds\cite{TLGR11}.

We can also use this approach instead of the tomographic approach described above to obtain a security bound for the rfi protocols.  That is possible by using the complete knowledge of the eigenvalues $\boldsymbol{\lambda}$ to infer the max-entropy in the virtual bases corresponding to the local rotations that optimized Alice and Bob's correlations.

The UR lower bounds Eve's uncertainty on the key,
\begin{equation}
H_{\text{min}}^{\bar{\epsilon}}(\mathbf{Z}|E) + H_{\text{max}}^{\bar{\epsilon}}(\mathbf{X}|B) \geq \log \frac{1}{c},\label{eq:UR}
\end{equation}
where $c$ quantifies the `incompatibility' between the measurements $\mathbf{Z}=Z^{\otimes n}$ and $\mathbf{X}=X^{\otimes n}$. Moreover, as any decent measure of uncertainty, $H_\text{max}$ can only increase under information processing and in particular under Bob's measurement~\cite{TCR10}, so
\begin{equation}
H_{\text{max}}^{\bar{\epsilon}}(\mathbf{X}|B) \leq H_{\text{max}}^{\bar{\epsilon}}(\mathbf{X}|\mathbf{X'})
\end{equation}
where the measurement $\mathbf{X'}=X'^{\otimes n}$ is made on system $B$. Now comes the crucial insight for what follows: the protocol does not have to prescribe the \textit{actual} measurement of $\mathbf{X}$ and $\mathbf{X'}$. Given the observed parameters, we are free to imagine the measurement $\mathbf{X'}$ on Bob's side that makes his uncertainty as small as possible on a \textit{hypothetical} measurement $\mathbf{X}$.  The data required for this is easily found once the state is reconstructed from the tomographic data. 

Let $X$ and $X'$ be observables mutually unbiased to $Z$ with outcomes in dimension $d$ corresponding to measurements on Alice and Bob respectively, whence the right hand side of (\ref{eq:UR}) reduces to $n\log d$.  Using a bound for $H_{\text{max}}^{\bar{\epsilon}}$ found using an extension of the method of~\cite{TLGR11},
\begin{equation}
H^{\bar{\epsilon}}_{\text{max}}(\mathbf{X}|\mathbf{X'})  \leq n H(\mathbf{v}_{X\otimes X'}(\mu)),
\end{equation}
finally we have to maximize $H\left(\mathbf{v}_{X\otimes X'}(\mu)\right)$, where $\mathbf{v}_{X\otimes X'}(\mu)$ is the vector of probabilities of the different outcomes (the parameters $\mathbf{P}$) in this virtual basis, up to a finite sampling uncertainty $\mu$.   In this way, we have been conservative in giving Eve maximum information on the key, compatible with our observed data.  The finite key rate reads therefore
\begin{equation}
r_N \leq \frac{n}{N}\Big(\log d - H(\mathbf{v}_{X\otimes X'}(\mu)) - \text{leak}_{\text{EC}}/n - \frac{2}{n} \log\frac{1}{2 (\epsilon_{\text{PA}}-\bar{\epsilon})}\Big)\,.
\label{eq:fkskr}
\end{equation}
We can maximize this value over the trade-off between the signals devoted to the key and those devoted to parameter estimation, as well as over the choices of $\epsilon_{\text{PE}}$ and $\epsilon_{\text{PA}}$ compatible with $\epsilon_{\text{sec}}$ for given $\epsilon_{\text{EC}}$ and now $\bar{\epsilon}$ is a function of $\epsilon_{\text{PE}}$.

\begin{figure}[h!]
\begin{center}
\includegraphics[width=0.8\textwidth]{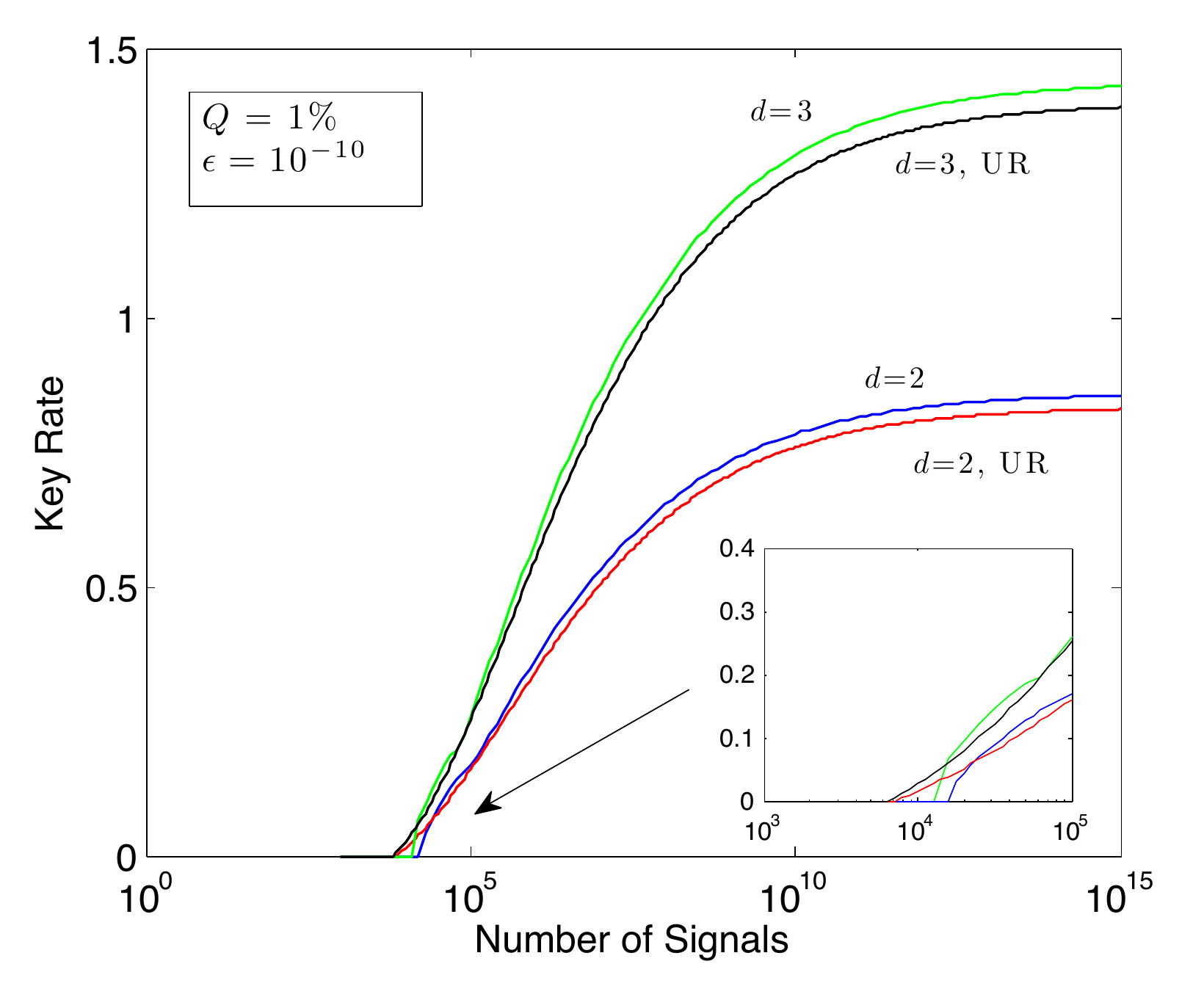} 
\caption{Finite key rate $r_{N}$ as a function of the number of signals $N$ for the dimensions $d=2$ and $3$ computed via uncertainty relation using equation ($\ref{eq:fkskr}$) and tomographic approach using equation ($\ref{eq:rn}$). The plots are for $\epsilon_{\text{sec}}=10^{-10}$, $\epsilon_{\text{EC}}=10^{-20}$ and $Q=1\%$.}
\label{fig:finite}
\end{center}
\end{figure}

\section{Conclusions} The derivation of the UR is generic: it is not linked to BB84, nor to QKD for that matter. Still, one may surmise that, in the context of QKD, such a UR is of practical use only for two-measurement protocols like BB84 and its higher-dimensional generalizations \cite{CBKG02,SS10}. When it comes to the six-state protocol, which uses three measurements, the UR yields the same bound as for BB84, while a better bound is found by taking into account the detailed structure of the states \cite{review2}.  In fact, in considering the rfi protocol, for which the bound described in the previous section with $d=2$ is equivalent to the bound for the six-state protocol, the asymptotic bound for the secret key is indeed worse using the UR technique.  However, the critical number of signals at which the secret key rate becomes positive is reduced using the UR approach (see Figure~\ref{fig:finite}).  This suggests that there should exist a better technique for constructing the finite-key bounds that gives the best of both methods, probably by eliminating the need for the smoothing correction term given in section 3.3.4 of Renner's thesis\cite{rennerthesis}.

\section*{Acknowledgments}

This work was supported by the National Research Foundation and the Ministry of Education, Singapore.  We would like to thank an anonymous referee for insightful feedback, and Markus Grassl and Stephanie Wehner for helpful discussions.

\section*{Appendix.}
The essential ingredients in the proof that equation~(\ref{eq:Cd}) generalizes $C$ are twofold: (i) the relation between average values of operators and the Hilbert-Schmidt inner product, namely $\langle O\rangle_\rho=\langle\rho,O\rangle$ where $\langle \cdot,\cdot\rangle$ is the Hilbert-Schmidt inner product defined as $\langle A,B\rangle=\Tr(A^\dagger B)$, and (ii) the Weyl operators as an orthonormal basis up to normalization, i.e.\ $\langle X^{k_1}Z^{\ell_1},X^{k_2}Z^{\ell_2}\rangle=d\delta_{k_1,k_2}\delta_{\ell_1,\ell_2}$. We recall the computation of inner product using an orthonormal basis
\begin{equation}
\langle\rho,\rho\rangle = \frac{1}{d}\mathop{\sum_{k,\ell=0}^{d-1}} \langle\rho,X^kZ^\ell\rangle\langle X^kZ^\ell,\rho\rangle = \frac{1}{d}\mathop{\sum_{k,\ell=0}^{d-1}} |\langle\rho,X^kZ^\ell\rangle|^2.
\end{equation}

To prove that $C$ is invariant with respect to rotations around $Z$, first note that $C$ can be rewritten as
\begin{eqnarray}
C &=& \mathop{\sum_{k_1,k_2=0}^{d-1}}_{\ell_1,\ell_2=0} |\langle X_A^{k_1} Z^{\ell_1} \otimes X_B^{k_2} Z^{\ell_2} \rangle |^2 - \mathop{\sum_{k_2,\ell_2=0}^{d-1}}_{\ell_1=0} |\langle Z^{\ell_1} \otimes X_B^{k_2} Z^{\ell_2} \rangle |^2 \nonumber \\ 
    &-& \mathop{\sum_{k_1,\ell_1=0}^{d-1}}_{\ell_2=0} |\langle X_A^{k_1}Z^{\ell_1} \otimes Z^{\ell_2} \rangle |^2 + \mathop{\sum_{\ell_1,\ell_2=0}^{d-1}} |\langle Z^{\ell_1} \otimes Z^{\ell_2} \rangle |^2
\end{eqnarray}
where the first sum simplifies to $d^2\Tr(\rho_{AB}^2)$. We can switch bases from $X_B^{k_2}Z^{l_2}$ to $X^{k_2}Z^{l_2}$ since they are both bases for $\mathcal{L}(\CC^d)$, thus invariant, and similarly for the third sum.  The final term is obviously invariant with respect to $Z$ rotations. Therefore, we have proved that $C$ is independent of the local unitaries $U$ and $V$ commuting with Z.

To show that $C$ acts as an entanglement witness, consider the product state $\rho_{AB}=\sigma_A\otimes\sigma_B$ for which $C$ factorizes into
\begin{equation}
C = \mathop{\sum_{k_1=1}^{d-1}}_{\ell_1=0} |\langle X_A^{k_1} Z^{\ell_1} \rangle_{\sigma_A} |^2 \mathop{\sum_{k_2=1}^{d-1}}_{\ell_2=0} |\langle X_B^{k_2} Z^{\ell_2} \rangle_{\sigma_B} |^2
\end{equation}
and note that
\begin{equation}
\mathop{\sum_{k_1=1}^{d-1}}_{\ell_1=0} |\langle X_A^{k_1} Z^{\ell_1} \rangle_{\sigma_A} |^2 = d\Tr(\sigma_A^2)-1-\mathop{\sum_{\ell_1=1}^{d-1}} |\langle Z^{\ell_1} \rangle_{\sigma_A} |^2 \leq d-1,
\end{equation}
from which $C\leq(d-1)^2$ for all product states and moreover for all separable states by convexity.  Thus if $C>(d-1)^2$ for a particular state, then the state is entangled, however, the converse, that the state is separable for $C$ less than that value, is not implied.  Indeed entangled states can have $C<(d-1)^2$.

Note that $C$ is a sum over tensor products of operators that do not commute with $Z$, the raw key basis.  The maximum value of $C$ is only achieved for maximally entangled states.   The maximum value that can be obtained with a separable state is $(d-1)^2$, therefore there is a gap between separable states and maximally entangled states that scales linearly with $d$.

\bibliographystyle{plain}
\bibliography{QKDBib12}

\end{document}